# A time of flight secondary ion mass spectrometry based molecular distribution distinguishes healthy and osteoarthritic human cartilage

Berta Cillero Pastor[1,2], Gert Eijkel[1], Andras Kiss[1], Francisco J. Blanco García[2], Ron M.A. Heeren[1*].

1- Biomolecular Imaging Mass Spectrometry (BIMS) FOM Institute AMOLF. Amsterdam. The Netherlands.

2- Rheumatology Division. Proteomic Unit. Proteo-Red-ISCIII. INIBIC-Hospital Universitario A Coruña, A Coruña, Spain.

*Correspondence to:

Ron M.A. Heeren, Ph.D.

FOM-Institute AMOLF. Science Park 104. 1098 XG Amsterdam. The Netherlands.

Tel: +31-20-7547100

Fax: +31-20-7547290

Email: heeren@amolf.nl

ABSTRACT. Osteoarthritis (OA) is a pathology that ultimately causes joint destruction. The cartilage is one of the principal affected tissues. Alterations in the lipid mediators and an imbalance in the metabolism of cells that form the cartilage (chondrocytes), have been described as contributors to the OA development. In this study, we have studied the distribution of lipids and chemical elements in healthy and OA human cartilage. Time of flight secondary ion mass spectrometry (TOF-SIMS) allows us to study the spatial distribution of molecules at a high resolution on a tissue section. TOF-SIMS revealed a specific peak profile that distinguishes healthy from OA cartilages. The spatial distribution of cholesterol-related peaks exhibited a remarkable difference between healthy and OA cartilages. A distinctive co-localization of cholesterol and other lipids in the superficial area of the cartilage was found. A higher intensity of oleic acid and other fatty acids in the OA cartilages exhibited a similar localization. On the other hand, $CN^-$ was observed with a higher intensity in the healthy samples. Finally, we observed an accumulation of calcium and phosphate ions exclusively in areas surrounding the chondrocyte in OA tissues. To our knowledge, this is the first time that TOF-SIMS revealed combined changes in the molecular distribution in the OA human cartilage.

## INTRODUCTION

Osteoarthritis (OA) is one of the most prevalent rheumatic diseases affecting more than a half of the population above sixty years of age (1). The main characteristic of the disease is the progressive loss of cartilage due to the imbalance between synthesis and degradation of the cartilage matrix (2). However, its precise etiology is still far from understood, and results in the lack of adequate biomarkers for early diagnosis and prediction of eventual joint damage. For these reasons, no complete and efficient treatment exists: the available therapies primarily comprise joint replacement, analgesics and inflammatory drug administration (3). Recent studies suggest that OA is a disease resulting from multiple pathophysiological mechanisms in which local and systemic factors play a role (4). It has been suggested that OA is likely to be a systemic disease involving lipid metabolism, as generalized changes in many tissues of the joints have been observed including adiposity (5, 6). In epidemiological studies, obesity coincided with a high incidence of non load-bearing types of OA, such as hand OA, suggesting a systemic disorder (7, 8). In general, lipids like cholesterol have been recognized as key regulators of normal vertebrate embryogenesis. However, excessive accumulation of cholesterol is toxic for the cells. This accumulation is prevented by tight regulation of infflux and efflux pathways (9). The only cell type present in cartilage (chondrocyte) is responsible to maintain the tissue homeostasis. Disorders in the lipid metabolism but also an imbalance in ion pumps and ion regulation, are linked to cell death and to cartilage destruction (10, 11).

Different methods have been used in the past to study lipids and small molecules in biological tissues. Electromicroscopy provides information about lipids in relation to organelles. Histochemical staining with Oil Red O or Sudan dye is used to nonspecifically localize lipids, while lipid-specific fluorescence probes are available for cholesterol only. Chromatographic methods previously used for lipid identification in the cartilage include gas chromatography, thin layer chromatography and high-performance liquid chromatography (HPLC) (12, 13). However, these techniques require extraction steps that are time consuming, need large amounts of tissues and often result in lipid mobilization. Moreover, these techniques cannot reveal the heterogeneous spatial distribution of lipids in the biological tissues. Imaging mass spectrometry (IMS) is a technique used in mass spectrometry to visualize the spatial distribution of compounds, metabolites, peptides or proteins by their molecular masses. Among the imaging techniques, time of flight secondary ion mass spectrometry imaging (TOF-SIMS) affords molecule-specific local information without the incorporation of a dye, isotopes or labels. Static TOF-SIMS is a surface-sensitive method where the primary ion dose is low, providing molecular information of the sample surface with minor damage (14, 15). In TOF-SIMS, the primary beam desorbs and ionizes molecular species from the sample surface, causing ejection of a large number of secondary ions that can be separated by mass (16). Displaying the mass spectra collected from the sample, we can generate chemical images. Thus, it allows us to compare the local signal intensity of different molecules in a biological tissue section. The use of surface metallization can also enhance the desorption/ionization of membrane components such as lipids and sterols of tissues and cells (17). Several studies have been successful in identifying intracellular distributions of specific biological ions, such as sodium, calcium, and membrane lipid fragments using TOF-SIMS (18-20). Other groups have described a potassium ion localization by 3D imaging TOF-SIMS specifically in thyroid tumor cells (21). Applications of the technology to the biomedical field have been also exploited. For instance, molecular ions from phosphatidylcholine, phosphatidylglycerol and protein fragments discriminate groups with asthma and healthy controls (22). Human striated muscle samples, from male control and Duchenne Muscular Dystrophy-affected children, have been subjected to TOF-SIMS detecting a different fatty acid (FA) chains composition depending on the local damage (23). Although this is a promising and widely known method, only a few studies have employed the technology to analyze the cartilage composition (24, 25) but none of them describe and compare the molecular distribution of human control and diseased OA cartilage by SIMS.

In this work we have employed TOF-SIMS to study the differences in the composition of healthy and OA human cartilage aiming to contribute to the understanding of lipid and ion regulation in OA cartilage.

## EXPERIMENTAL SECTION

### Cartilage procurement and processing

Control human knee cartilage from adult donors with no history of joint disease was provided by the Tissue Bank and the Autopsy Service at CHU de La Coruña, Spain. Osteoarthritic (OA) cartilage was obtained from consenting donors who were undergoing joint replacement. Cartilage slices were removed from the condyles under aseptic conditions and frozen in liquid $N_2$. For tissue studies, pieces of 10 µm of thickness were cut in a cryostate and deposited on indium tin oxide (ITO) high conductivity slides (Delta Technologies, CO, USA) and frozen at -20ºC. This study was approved by the local ethics committee in Galicia, Spain.

### Metal deposition

Samples were kept in a desiccator box, defrosted and dried in a vacuum desiccator to prevent the molecular de-localization prior the metal deposition. Two 2 nm of gold was deposited on top of the samples using a Quorum Technologies (Newhaven East Sussex, UK) SC7640 sputter coated equipped with a FT707 quartz crystal microbalance stage to enable metal assisted secondary ion mass spectrometry (MetA-SIMS). A FT7690 film thickness monitor was used throughout the deposition process.

### TOF-SIMS experiments

TOF-SIMS experiments were performed in positive and negative modes on a Physical Electronics (Eden Prairie, MN, USA) TRIFTII secondary ion mass spectrometer with an $Au^{1+}$ primary ion beam. Experiments were performed with a primary ion beam of 7.2 nA, a primary pulse length of 19 ns and a dose of 4.47x10^11 ions/cm$^2$ on the sample surface. This value is below 10^13 ions/cm$^2$, also called static limit which means that measurement was terminated before the surface had been significantly damaged by the primary ion beam. The sample surface was observed continuously during the analysis using an integrated video camera. Areas of 2x2 mm were analyzed using the mosaic mode of 2569x256 pixels per tile. The raster size was set at 125 µm per tile and a lateral resolution of 488 nm. The acquisition time was 10 seconds per tile. The data acquisition was performed with the software WinCadence version 4.4.0.17 also used for data visualization (PhysicaL Electronics, Chanhassan, USA).

**Data interpretation**

The **supplementary figure 1** describes the data analyses workflow. Briefly, raw data was converted to Matlab format. Spatial binning, spectral binning and peak picking were performed prior selecting the 500 *m/z* channels with the highest variance. After selecting the regions of interest (ROI), a total of 10983 and 5727 spectra, acquired in positive and negative mode respectively, were used for the analysis. Principal component analysis (PCA) was used to look for spectral similarities and differences between the samples using our in-house built ChemomeTricks toolbox for MATLAB version 2009b (The MathWorks, Natick, MA). PCA extracted linear combinations of variables, each of which is associated with the largest possible variance after removing the variance of prior principal components in an iterative fashion. A specific category number was assigned to the spectra from each donor. The same category number was assigned to replicates from the same donor. Discriminant analysis (DA) was performed to look for the peaks with the highest differences between all the OA and healthy categories. PCA and DA described the specific spectra of different areas and distinguished between different conditions (healthy and OA). The scores that described the different Discriminant functions (DF) were adjusted to Gaussian curves following a normal distribution. The score distribution described the sample identity. After data analyses, *m/z* of interest were selected. The assignments of *m/z* identities were performed according to the bibliography and Lipimaps.org database (http://www.lipidmaps.org/).

**Lipid staining**

For lipid staining, samples were dipped in isopropyl alcohol 60% in a closed recipient. Samples were then stained with an oil red solution for 20 min (oil red 0.3 gr/L in isopropyl alcohol). After washing in water, they were stained with hematoxylin and mounted.

### Hematoxylin staining

The slides were immersed in hematoxylin (Sigma-Aldrich) for 1 min and then dipped in running warm tap water for 15 min for nuclei and cytoplasm staining. A wash in destillated water for 2 min followed by a wash in EtOH 95% (2 min) were performed before eosin staining (30 sec). Tissues were rinsed in deionized water (2 min), in EtOH 95% (2x 2 min) and in EtOH 100% (2 min). Digital images were acquired with the Mirax system (Carl Zeiss, Siledrecht, The Netherlands) after the dehydrating steps.

## RESULTS AND DISCUSSION

### TOF-SIMS analysis of human and OA cartilage revealed an OA mass profile

In the present work, we have acquired the mass spectra from 0 to 2000 Da of three healthy and three OA human cartilage donors in duplicate. After PCA and DA two different groups corresponding to healthy and OA samples were clearly distinguished in the positive ion mode. The highest variance, represented in the first Discriminant function (DF1), separated healthy and OA samples. Plotting the scores of DF1 versus the frequency of occurrence of these scores, two different distributions (healthy and OA) with a Gaussian shape were observed (**figure 1A**). DA revealed OA specific peaks according to the positive loadings of DF1. The masses that were specific to the OA samples were *m/z* 339.24, *m/z* 351.00, *m/z* 368.99, *m/z* 370.95, *m/z* 381.02 and *m/z* 383.26 (**figure 1B)**. According to the tentative assignments using the Lipidmaps database and a lipid MS literature search, these peaks are tentatively attributed to different fragments of vitamin D-3 (*m/z* 383.3), prostaglandins and glycerolipids (*m/z* 351.00, *m/z* 368.99, *m/z* 370.95, *m/z* 381.02). In addition to this, the peak *m/z* 339.1 has been found in macrophages stimulated with interferon gamma (IFN-γ) and lipopolysaccharide (LPS) and other authors have described it as a a monoacylglycerol MAG (C18:1) (26, 27). Interestingly, IFN is an inflammatory mediator related to many rheumatic diseases and that regulates the production of prostaglandins in cartilage and synovial tissue. Further analyses including MSMS-based determination of the assigned peaks are required to confirm this hypothesis but are beyond the current capabilities of the ToF-SIMS instruments currently available to us.

Data normalization and the projection of DF1 scores were performed to study the distribution of OA and healthy specific masses in the two types of samples. The reconstruction of these images, represented a higher average intensity of the negative scores (DF1-) in the healthy samples (**figure 2A**) and a higher average intensity of the positive scores (DF1+) in the OA samples (**figure 2B)**. In addition, the projection of the positive scores (OA specific), revealed a distribution in the superficial area of the healthy tissue (next to the synovial membrane tissue). In the case of OA cartilage the DF1 positive projections, showed intense droplets, with a higher intensity in the superficial area compared to the deep area (next to the subchondral bone). Thus, by positive TOF-SIMS and DA we could distinguish between healthy and OA cartilages according to a group of masses which were specific to each group. In addition, we were able to localize these aberrant molecular signatures to specific areas of the cartilage. More analyses will be performed in the future to study the possibility of discriminate between different OA grades according to the molecular signatures found. The distribution of OA peaks seemed to be localized in the superficial area of the cartilage. This area is in contact with synovial fluid, an important source of inflammatory factors extensively described in the literature (2, 28).

**PCA revealed the presence of droplets rich in lipids in OA cartilage**

PCA was used to analyze the intra-tissue differences in the spectral composition in healthy and OA cartilages independently. In the healthy samples, the first function separated the extracellular matrix (ECM) (represented by the positive part +1) from the cell membranes (represented by the negative part -1) **(figure 3A**). The spectra that described the first PCA function, showed that cell membranes are rich in cholesterol and related species (**figure 3B**). Examples are the dominant cholesterol water-loss fragment $[M\text{-}OH]^+$ at *m/z* 369, the pseudo-molecular ion $[M\text{-}H]^+$ at *m/z* 385 and *m/z* 970, corresponding to the $[2M+Au]^+$ ion. Peaks related to the Diacylglycerols (DAG) can be also detected in the same principal component, as exemplified by peaks at *m/z* 577 (C34:1) and 603 (C36:2) respectively. The second PCA function separated the spectra of the lacunae (space where one or more cells are embedded) (+2) from the intracellular material (-2). In the OA tissue the differences between the cell membranes and the ECM spectra were not represented in the first function. In OA cartilage, the cholesterol species like the dominant cholesterol fragments $[M\text{-}OH]^+$ at *m/z* 369 or *m/z* 385, *m/z* 970 and DAG (C24:1), were accumulated in droplets in the superficial area (+), and clearly separated from the deep ECM **(figure 3A, 3C).** The second function separated the ECM from the superficial and the deep area of the cartilage. The third function separated the lipid droplets in the superficial area, from the pericellular space. Thus, TOF-SIMS and PCA statistical analyses, revealed the intra-sample heterogeneity in healthy and OA human cartilage. In addition, TOF-SIMS revealed an accumulation of lipids exclusively in the superficial area of OA cartilages demonstrating the importance of the lipid metabolism alteration in the OA pathology.

Due to the fact that the OA specific droplets were rich in lipids and negative ion mode is extremely suited to study fatty acids and other lipids, human healthy and OA cartilages were also analyzed in the negative mode. After DA the spectra of healthy and OA tissues were separated in two groups.  In addition, DF1 scores had a different distribution in healthy and OA samples, both following a Gaussian curve (**figure 4A**). This means that we are able to classify different types of cartilage samples. The first DA function revealed specific masses from OA cartilages (positive loadings) and healthy cartilage (negative loadings). Some examples of OA specific peaks were *m/z* 281.2 related to C18:1 (oleic acid) (DF loading=0.87) and other masses like *m/z* 255.2 from C16:0, (palmitic acid; DF loading=0.83), *m/z* 253.2 from C16:1 (palmitoleic acid; DF loading=0.81), *m/z* 283.1 from C18:0 (stearic acid; DF loading=0.73) and *m/z* 279.2 from C18:2 (linoleic acid; DF loading=0.76). The high positive loadings for these masses represented their specificity to OA samples. Other peaks like Vitamin E/alpha, *m/z* 429.2, were also increased in lipid rich areas of the OA samples (29). On the other hand, the mass *m/z* 26.3 ($CN^-$) was specific from healthy samples (30). $CN^-$ is a marker of protein content and it has been associated to protein rich tissues areas (31). Cell death and tissue destruction are typical phenomena of the OA disease. In addition, matrix degradation products generated by excessive proteolysis in arthritic pathologies have been found to mediate cartilage destruction, explaining the lower protein abundance in OA tissues. The decreased *m/z* 26 intensity in OA cartilage is supported by this theory.

The study of the raw data acquired after TOF-SIMS tissue experiments already showed the differences in ion intensity in some of the masses of interest, confirming the results obtained after data reduction and interpretation. **Figure 5A** exemplifies the average raw spectra over the entire section of a healthy and an OA sample (*m/z* range of 200-400) measured in the negative mode. A higher intensity of *m/z* 255.2 and 281.2 was observed in OA tissues. **Figure 5B** represents a spectrum in the mass range of 0-200 *m/z* in the negative mode revealing and displaying a higher intensity of *m/z* 26 in the healthy samples.

As explained above PCA and DA were employed to study the distribution and the intensities of several groups of masses. We have also studied by univariate analyses the distribution of OA specific masses (*m/z* 253.2, *m/z* 255.2, m/ 279.2, *m/z* 281.2, *m/z* 283.1 and *m/z* 429.2) (**figure 6).** With the exception of palmitoleic acid (*m/z* 253.2) that exhibited a more general distribution, all the masses were located in the cholesterol rich areas. According to Leistad et al., a liberation of oleic acid can be produced when chondrocytes are stimulated with tumor necrosis factor (TNF), one of the main mediators of inflammatory processes (32). An important number of masses in the mass range of phospholipids (600-900) have been found in the OA samples according to the DA. The FA detected and here described as masses specific to the OA condition, could be part of these lipids. Future parallel studies using MALDI-imaging for instance, could discriminate these intact lipids as other authors have shown (33-35).

Alteration in cholesterol and lipid content have been associated in many diseases related to age like obesity or diabetes (30). In neurodegenerative diseases lipid anomalies potentially linked to Alzheimer Disease (AD) have been also reported by mass spectrometry (36) and, in cardiovascular pathogenesis, the accumulation of cholesterol is one of the primary risk factors (37, 38). In the case of the cartilage, only few authors have studied the lipid metabolism of chondrocyte cells. A relation between cartilage lesion and alterations in the lipid metabolism has been found by techniques such as electron microscopy (39, 40). Finally, FA, DAG and other lipids have been detected in high concentrations in plasma of OA patients by LC-chromatography (41) supporting our results.

**Validation of lipid localization**

As we explained above, the specific deposition of lipid precipitates in the OA samples was an interesting feature that suggested the metabolic alteration of OA cartilage cells. The local environment can have an effect in the ionization of specific molecules. Since the ECM can vary in composition between different areas of the cartilage and this could affect the ionization, it is important to validate the results we found with other techniques to make solid conclusions. We validated the content of these lipid droplets by oil red staining. **Figure 7** shows the lipid distribution in healthy and OA cartilage. In OA samples big lipid droplets (red) were observed in the superficial area of the cartilage, confirming the TOF-SIMS results. The oil red stain merely confirms the compound class present. Unlike TOF-SIMS it provides no hints towards the actual chemical composition and acyl chain length distribution. The fact that all these lipid related species were localized in droplets in the superficial area, revealed the important role of the adjacent tissues (synovial membrane) and fluids (synovial fluid) in the process of inflammation and destruction of the cartilage. The pathways involved in the cross-talk between cartilage and other tissues are not completely understood. Here we have revealed a clear pathological role of lipid related molecular pathways in the OA pathology. We therefore conclude that an alteration in the lipid metabolism could be a risk factor and/or a consequence of the OA disease that should be studied in detail.

**Different localization of elemental ions**

Defects in ion pumps, homeostasis ion regulation and cell death are other characteristics of OA chondrocytes (42). Thus, alterations in the electrolyte balance can cause changes in the cartilage. The visualization of masses corresponding to elementary molecules reported interesting information. The study of the localization and the intensity of sodium ions were similar in both types of samples (**supplementary figure 2**). However the calcium signal (*m/z* 40) was distributed in deposits in the ECM. Other species measured in the negative mode, like the chlorine (*m/z* 35) had a similar distribution although a slightly lower intensity in OA than in healthy samples. We confirmed also a lower intensity of $CN^-$ in OA samples, especially in the areas with high content of lipids. We have also observed deposits of phosphates $PO^{2-}$ (*m/z* 63.1) and $PO^{3-}$ (*m/z* 79.2) only in OA cartilages. The calcification of the cartilage is another consequence of the OA pathology. The detection of crystals rich in calcium phosphate in synovial liquid from patients with OA and an increase in calcium and phosphorus levels in the OA cartilage, have been described in the literature (43, 44). All these facts contribute to the production of prostaglandins, other inflammatory factors and to change the biomechanical properties of the cartilage.

## CONCLUSIONS

In this study we investigated the possibility of using TOF–SIMS as a new method for the detection of changes in lipid and elemental composition in healthy and OA human cartilage. We were able to describe an OA specific peak profile. Accumulation of cholesterol and FA was detected in the superficial area of the OA cartilage. A disease specific accumulation of calcium and phosphates was observed only in OA cartilage. These results demonstrate that TOF-SIMS is a useful tool in the study of lipid and elemental composition changes associated with rheumatic pathologies with excellent diagnostic potential.

**ACKNOWLEDGEMENTS**

This work is part of the research program of the Stichting voor Fundamenteel Onderzoek der Materie (FOM), which is financially supported by the Nederlandse organisatie voor Wetenschappelijk Onderzoek (NWO).

**Competing interests:** The authors declare that they have no competing interests

## FIGURES

## Figure 1

A)

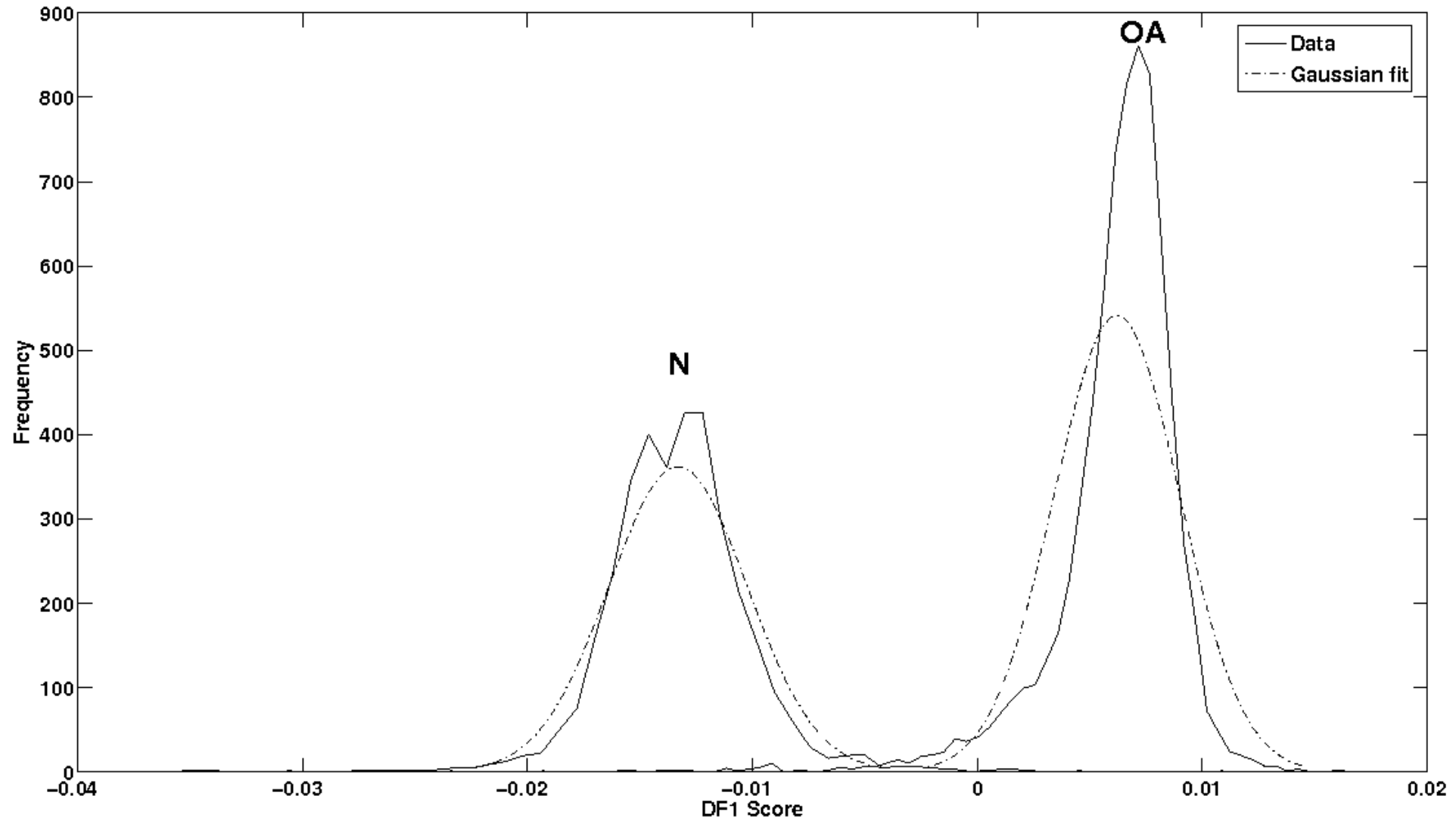

B)

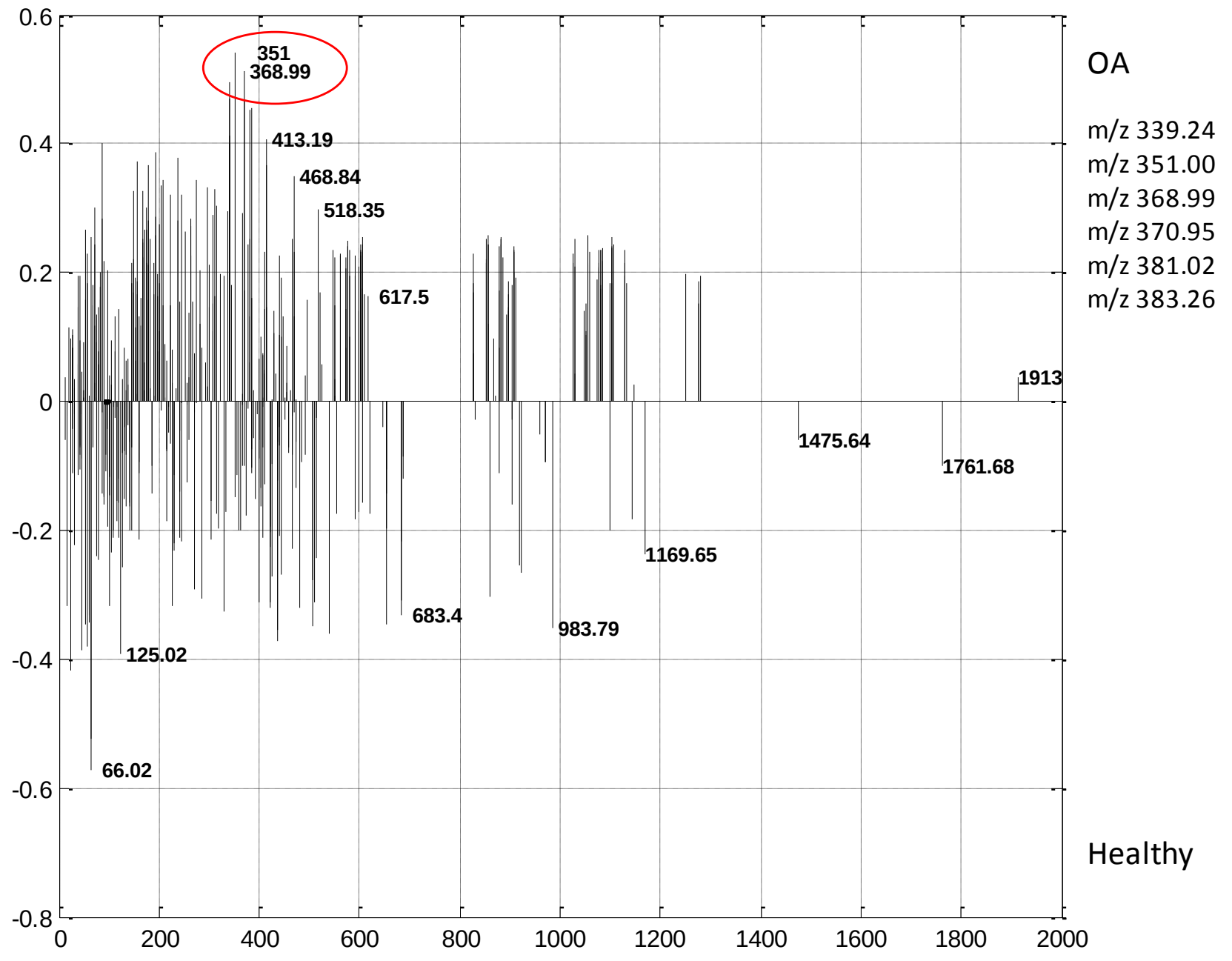

**Figure 1. Discriminant analyses revealed a different mass spectrum in healthy and OA human cartilage after positive TOF-SIMS.** A) Histogram distribution of DF1

scores where negative scores were specific to healthy samples and positive scores to OA samples. B) Spectra plot of the DF1. The list of OA specific peaks is shown.

**Figure 2**

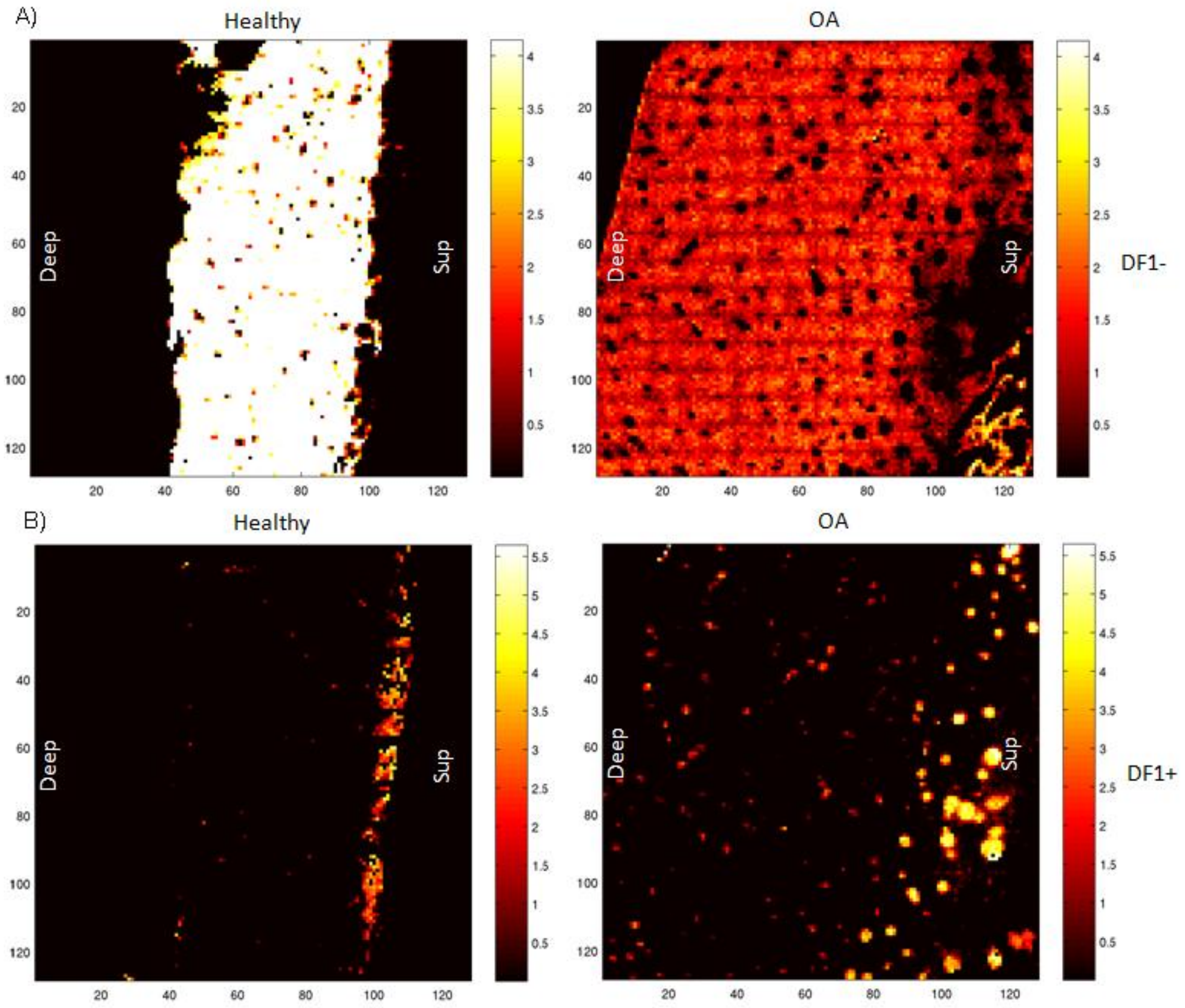

**Figure 2. First discriminant function (DF1) score plot projections.** The distribution of healthy (DF1-) and OA (DF1+) scores was studied. Superficial and deep areas are labeled. A) DF1 negative scores projection. The intensity of DF1 negative scores was higher in healthy than in OA cartilages. X and y axes represent the pixel number. B) The intensity of DF1 positive scores was higher in OA compared to healthy cartilages. The distribution was specific to the superficial area (next to the synovial membrane), in the case of OA cartilage in small droplets. X and y axes represent the pixel number.

**Figure 3**

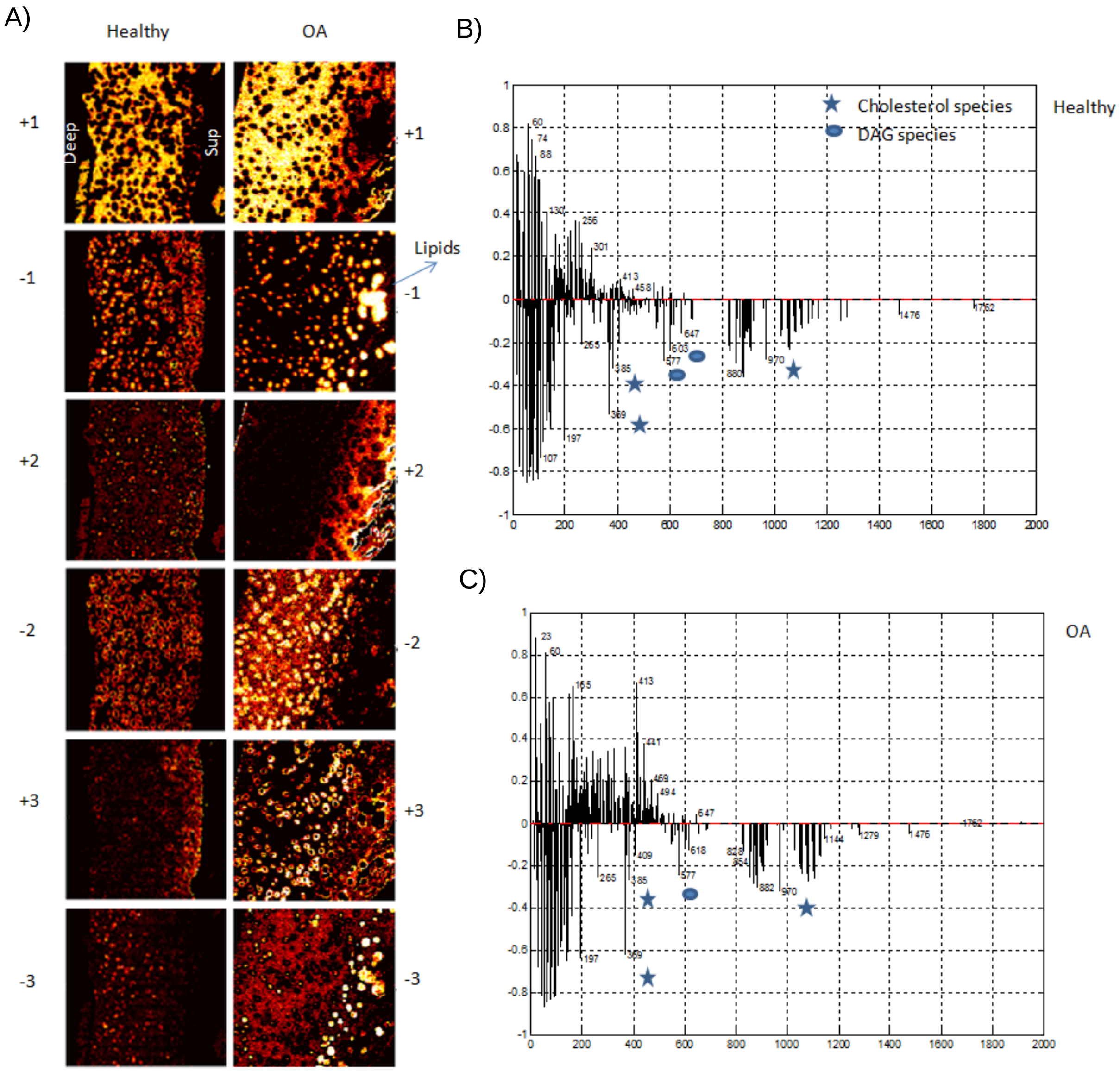

**Figure 3**. **Principal component analyses of healthy and OA human cartilage revealed intra-sample heterogeneity.** Superficial and deep areas are labeled. A) First three PCA functions in healthy and OA cartilages (positive mode). B) Spectra of the first PCA function in healthy cartilage that shows the lipid content of the cellular membranes (-1). Cholesterol (star) and DAG masses (circle) are shown. C) Spectra of the first PCA function in OA cartilage. Cholesterol and DAG masses are shown. Cholesterol and lipid droplets can be appreciated in the superficial area, exclusively in OA samples.

**Figure 4**

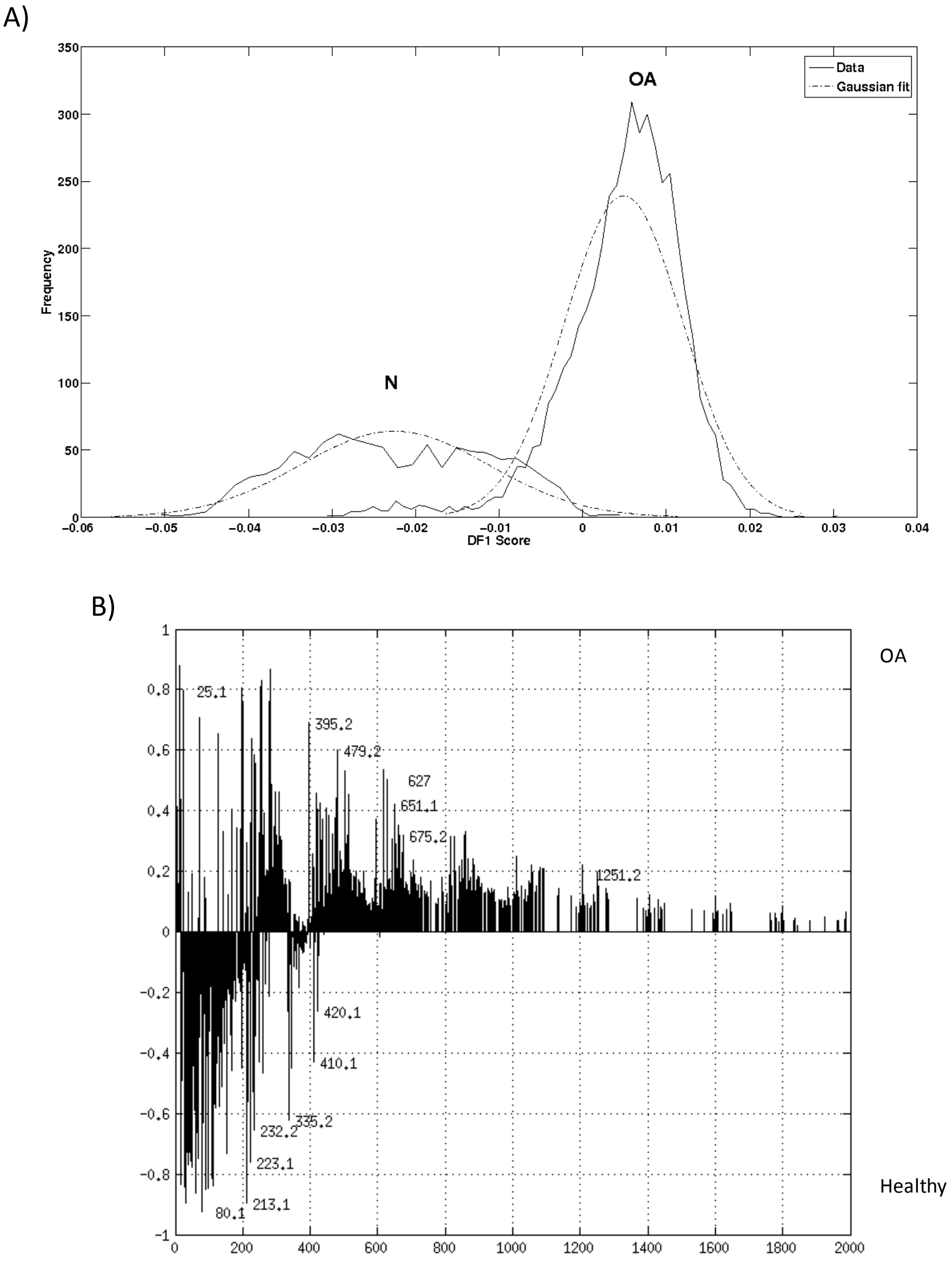

**Figure 4. Discriminant analyses revealed a different spectrum in healthy and OA human cartilage in the negative mode.** A) Histogram distribution of DF1 scores where negative scores were specific to healthy samples and positive scores to OA samples. B)

Spectra plot of the DF1. Fatty acids (*m/z* 253.2, *m/z* 255.2, m/ 279.2, *m/z* 281.2, *m/z* 283.1) were more abundant in OA cartilage.

**Figure 5**

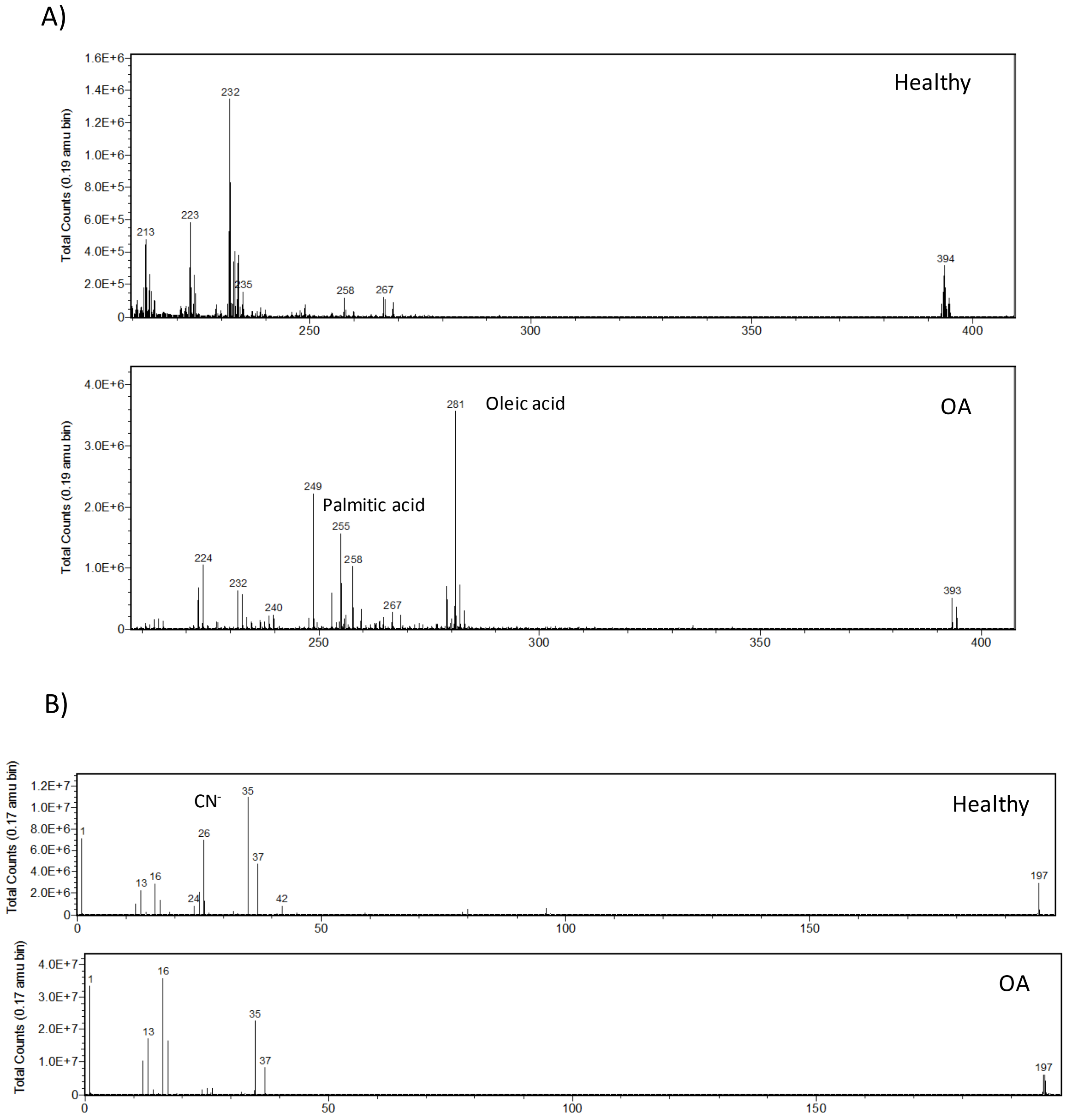

**Figure 5. Raw spectra of A) 200-400 *m/z* and B) 0-200 *m/z* acquired in the negative mode.** The figure shows the differences in the intensity of selected masses like palmitic acid (*m/z* 255.2), oleic acid (*m/z* 281.2) and CN- (*m/z* 26).

## Figure 6

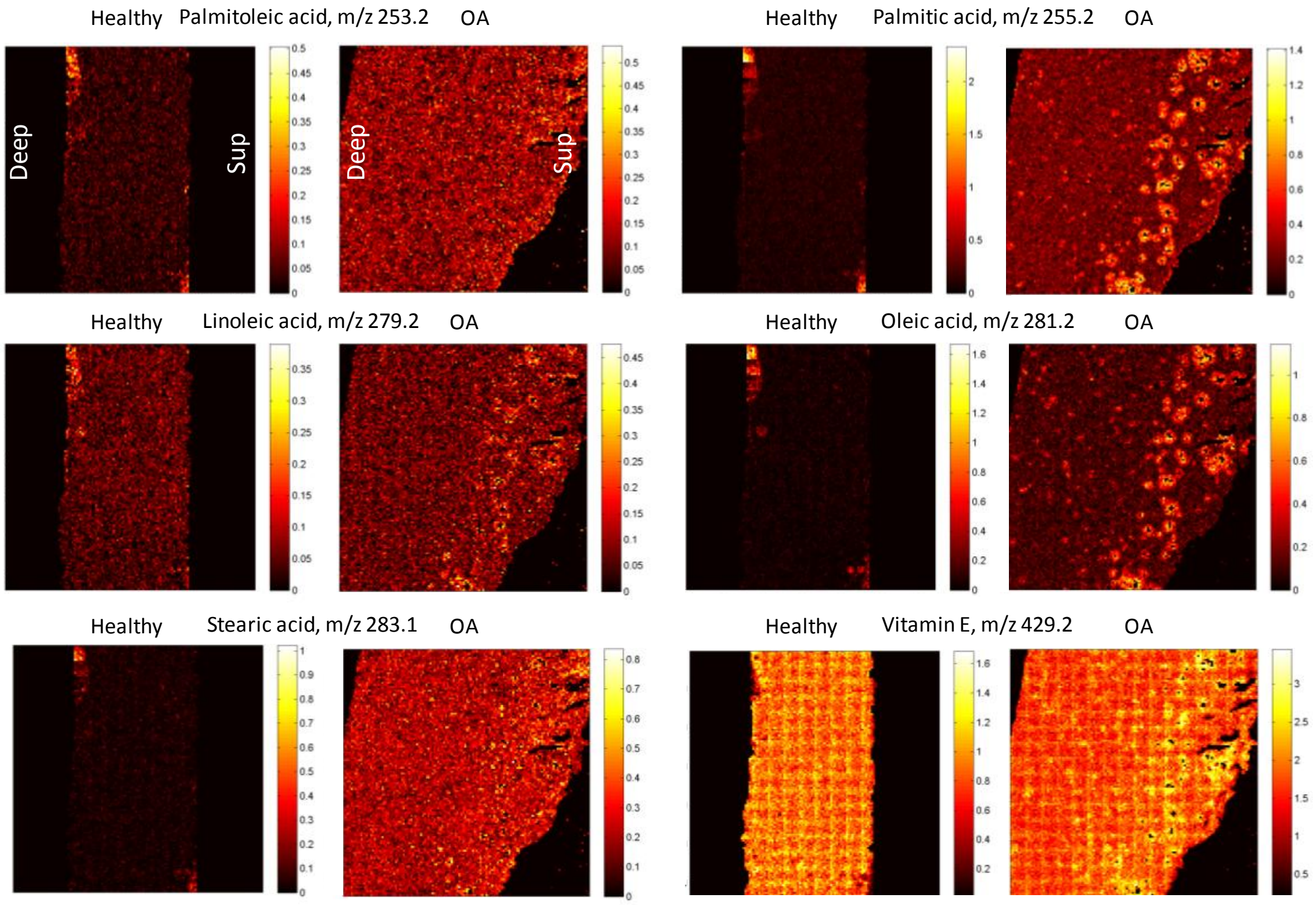

**Figure 6. Intensity and distribution of OA specific masses detected in the negative mode.** Different masses *m/z* 255.2, m/ 279.2, *m/z* 281.2, *m/z* 283.1 and *m/z* 426.2 were specifically localized in the superficial area of the OA cartilage (next to the synovial membrane).

**Figure 7**

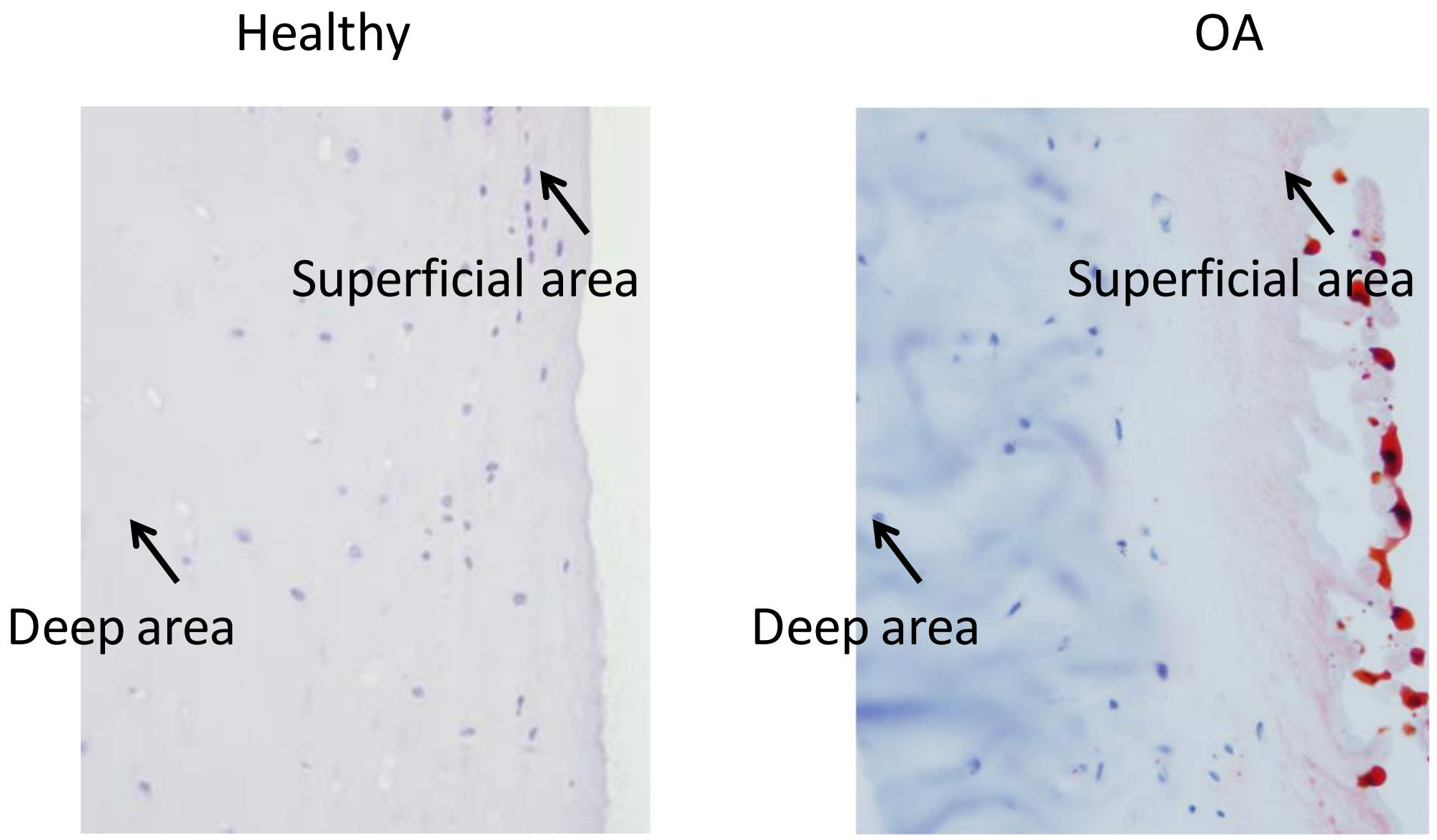

**Figure 7. Oil red/hematoxylin staining in healthy and OA cartilages validated TOF-SIMS experiments.** Lipid droplets can be appreciated in the superficial area of the OA cartilage.

**For TOC only**

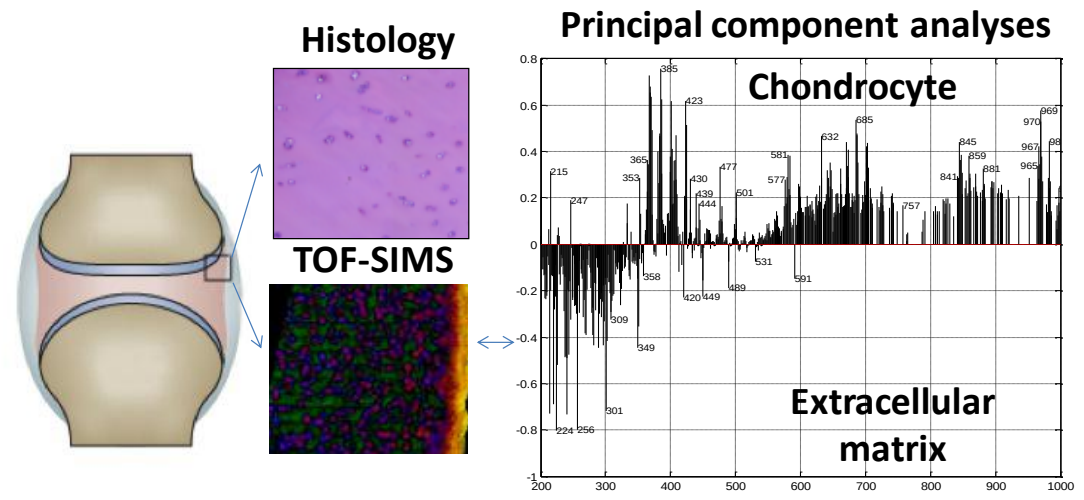

**Abstract graphic. TOF-SIMS and the study of the cartilage composition.** The figure shows the representation of the human joint and cartilage with a schematic of a human joint and a hematoxylin-eosin staining of cartilage. The figure also shows an overlay of the third (+3/-3) and the first PCA functions (+1) in a healthy cartilage after TOF-SIMS measurements. Each color represents a specific spectrum, distinguishing the chondrocyte cells (pink color), the lacunae (blue color) and the extracellular matrix (green) among other areas. The spectra of first PCA function is shown, revealing the specific composition of cells (rich in cholesterol and other lipids).

**Supporting Information Available:** This material is available free of charge via the Internet at **http://pubs.acs.org.”**